\documentclass[aps,prd,showpacs,onecolumn,preprintnumbers,amsmath,amssymb,nofootinbib,12pt]{revtex4-2}
\usepackage{graphicx} 

\input psfig.sty

\newcommand{\bq}{\begin{equation}}
\newcommand{\eq}{\end{equation}}
\newcommand{\bqn}{\begin{eqnarray}}
\newcommand{\eqn}{\end{eqnarray}}

\newcommand{\lb}{\label}

\def\gappr{\lower 3pt\hbox{$\buildrel > \over \sim\;$}}
\def\gappl{\lower 3pt\hbox{$\buildrel < \over \sim\;$}}
\def\limiter{\lower 7pt\hbox{$\buildrel{\textstyle\longrightarrow}\over{\scriptscriptstyle ~~s\rightarrow\infty~~}\;$}}
\def\ablim{\lower 9pt\hbox{$\buildrel{\textstyle\longrightarrow}\over{\scriptscriptstyle ~~~a\rightarrow b~~~}\;$}}
\def\x0lim{\lower 11pt\hbox{$\buildrel{\textstyle\longrightarrow}\over{\scriptscriptstyle ~~x^0\rightarrow-\infty~~}\;$}}
\def\xlim{\lower 8.5pt\hbox{$\buildrel{\textstyle\longrightarrow}\over{\scriptscriptstyle ~~x_\pm\rightarrow-\infty~~}\;$}}
\def\T0lim{\lower 11pt\hbox{$\buildrel{\textstyle\longrightarrow}\over{\scriptscriptstyle ~~T\rightarrow0~~}\;$}}
\def\Tlim{\lower 8.5pt\hbox{$\buildrel{\textstyle\longrightarrow}\over{\scriptscriptstyle~~T\rightarrow\infty~~}\;$}}
\def\Tmg1{\lower 8.5pt\hbox{$\buildrel{\textstyle\longrightarrow}\over{\scriptscriptstyle~~T>>1~~}\;$}}
\def\pdot{\raise 1.5pt\hbox{.}}

\def\dal{\hbox{$\sqcup$\hbox to 0pt{\hss$\sqcap$}}}

         \def\ln{{\rm ln}}
         \def\ex{{\rm e}}
         \def\dal{\hbox{$\sqcup$\hbox to 0pt{\hss$\sqcap$}}}

\begin{document}
            


\title{\Large Low mass strange stars and the compact star 1E\,1207.4-5209  
in the Field Correlator Method}
\author{F. I. M. Pereira}
\email{flavio@on.br, fimpjm@gmail.com} 
\affiliation{Observat\'orio Nacional, Rua Gal. Jos\'e Cristino 77, 
20921-400 Rio de Janeiro RJ, Brazil}
\date{\today}

\begin{abstract}
 
 We investigate the possible existence of anomalous mass defects in 
the low mass region of stellar sequences of strange stars.  
 We employ the nonperturbative equation of state derived in the 
framework of the Field Correlator Method to describe the hydrostatic 
equilibrium of the strange matter. 
 The large distance static $Q{\bar Q}$ potential $V_1$ and the gluon 
condensate $G_2$ are the main parameters of the model.

 We use the surface gravitational redshift measurements as a probe to 
determine the ratio $({\cal P}/{\cal E})_C$ at the center of strange stars.
 For $V_1=0$ and $G_2\gappr0.035\,{\rm GeV}^4$\,,  we show that 
$({\cal P}/{\cal E})_C\simeq0.262$ and the corresponding redshift 
$z_S\simeq0.47$ are limiting values, at the maximum mass of the highest 
mass stellar sequence. 
 As a direct application of our study, we try to determine the values of 
$V_1$ and $G_2$  from astrophysical observations of the compact star  
1E\,1207.4-5209.  
Due to the uncertainties in the surface redshift determination, we made two 
attempts to obtain the model parameters. 
 Our findings show that $({\cal P}/{\cal E})_C=0.073^{+0.029}_{+0.024}$\, at 
68\% confidence, $V_1=0.44\pm0.10$\,GeV at 90\% confidence and 
$G_2=0.008\pm0.001\,{\rm GeV}^4$ at 95\% confidence \, in the first attempt; and 
$({\cal P}/{\cal E})_C=0.073^{+0.029}_{+0.024}=0.087\pm0.028$\, at 71\% confidence, 
$V_1=0.43\pm0.085$\,GeV at 94\% confidence and 
$G_2=0.0093\pm0.00092\,{\rm GeV}^4$ at 94\% confidence in the second attempt. 
 These values of $V_1$ and $G_2$ are in reasonable agreement with the lattice and 
QCD sum rules calculations.
 As a consequence of the high values of $V_1$ and $G_2$, the anomalous mass defects of 
1E\,1207.4-5209 are $|\Delta_2M|\simeq2.56\times10^{53}$\,erg\, in the first attempt and $|\Delta_2M|\simeq2.94\times10^{53}$\,erg\, in the second attempt. 

\end{abstract}

\pacs{04.40.Dg, 21.65.Qr, 21.65.Mn} 

\maketitle


\section{\bf Introduction}
\lb{int}

 The possibility of anomalous mass defects in compact stars goes back to the 
works of V. A. Ambartsumyan, G. S. Saakyan and Yu. L. Vartanyam in  
Refs.\,\cite{Amb1,Amb2,Amb3,Sak,Amb5}. 
 Anomalous mass defects would occurs at internal stellar energy densities many 
times greater than the nuclear density ($\rho_0\simeq2.5\times10^{14}\,{\rm g\,cm^{-3}}$).
 Such stellar configurations, in the presence of external perturbations, would 
undergo transitions of explosive character, from a metastable state to a stable 
state, with great amounts of liberated energy. 
 The authors of Refs.\,\cite{Amb1,Amb2,Amb3,Sak,Amb5} considered the superdense 
stellar matter made of a degenerate gas comprising neutron, protons, hyperons and 
electrons, at zero temperature.

 As baryons are made of quarks, it would be natural to expect unbound quarks to 
exist in the interior of hyperdense stars.
 The possibility of hypothetical compact stars made of pure quark matter was  
then considered by N. Itoh in Ref.\,\cite{Ito}.  
 Since the Bodmer-Terazawa-Witten conjecture in Refs.\,\cite{Bod,Wit,Ter} the 
existence of the strange quark matter (SQM), made of an equal number of up, down 
and strange quarks, has been subject of a lot of theoretical studies, 
experimental investigations in terrestrial laboratories, and in observational 
studies of astrophysical phenomena.
 
 The properties of the SQM in the phase diagram, at small temperatures and large densities, 
were not completely known due to the nonperturbative character of quantum chromodynamics 
(QCD). 
 Within this scenario the Nambu-Jona-Lasinio Model in Refs.\,\cite{NJL1,NJL2} and 
the MIT Bag Model in Ref.\,\cite{MIT}  appeared to describe the properties of quark matter.
 The Nambu-Jona-Lasinio Model has been used to investigate quark matter properties in 
compact stars, as in Refs.\,\cite{DPM1,DPM2}. 
 It exhibits chiral symmetry breaking, but it has the disadvantage in that neither the quark 
confinement is explicitly included nor the contribution of gluons to account for the dynamics 
of the quark confinement is considered.

 The MIT Bag Model became one of the most used phenomenological models 
of quark confinement to describe the cold SQM at finite chemical potentials and to investigate 
the properties of strange stars, as in Refs.\,\cite{Far,AFO,HZS,Koh}. 
 However, the model also has the disadvantage in that the quarks are free particles inside the 
bag that simulates the confinement. 
 For larger distances, when the confining forces become important, it does not account for 
the way the quarks conglomerate to form hadrons in the quark-hadron phase transition. 
 In other words, the model does not take into account the $Q{\bar Q}$ interaction potential  
and the role of the gluon condensate to describe the quark confinement.
 Summarizing, neither Nambu-Jona-Lasinio Model nor MIT Bag Model naturally include the dynamics 
of quark confinement from the first principles of QCD.

 In another interesting approach, the quark interaction via the Richardson potential 
(see Ref.\,\cite{Ric}), which incorporates the aspects of the asymptotic freedom and quark 
confinement, has been used to study the properties of strange stars in Ref.\,\cite{Dey} and to 
describe the SQM in the presence of magnetic field in Ref.\,\cite{SHS}.

 Despite the highly nonlinear character of QCD, the inherent difficulties were overcoming by 
the advent of the nonperturbative equation of state derived, from first principles, in the 
framework of the Field Correlator Method (FCM) in Ref.\,\cite{Si6}.
 The great advantage of the FCM approach is that it covers the entire phase diagram plane from 
high temperatures and low densities to low temperatures and high densities. 
 We have made applications of the FCM nonperturbative equation of state to investigate general 
aspects of strange stars in Ref.\,\cite{Fla1} and the SQM stability in Ref.\,\cite{Fla2}.
 The authors of Ref.\,\cite{LoB} have made an application of the method to investigate quark 
deconfinement transition in neutron stars.
 More recently,  we have also considered the (normal and anomalous) mass defects of strange 
stars at the maximum masses of the stellar sequences\,, in Ref.\,\cite{Fla3}\,.

 In the present article, we study the anomalous mass defects of nonrotating strange stars 
without crust (or bare strange stars) and without internal magnetic field, within the same 
lines of our previous investigation in Ref.\,\cite{Fla3}. 
 The mass-radius relations of strange stars with crust, which have been investigated within 
the MIT Bag Model, are similar to that of neutron stars, as shown in Fig.\,8.5 in 
Ref.\,\cite{HPY}. 
 In the FCM, a richer approach with two parameters, the inclusion of a crust requires a 
more detailed investigation to be considered in future works. 
 In other words, depending on the values of the model parameters, it is not certain that 
the results obtained with the FCM will present the same features as the ones obtained with 
the MIT Bag Model.
 It may be premature the inclusion of a crust because it could mask the values of the 
FCM model parameters.
 From now on, strange stars means bare strange stars. 
 
 The nonperturbative treatment of the quark-hadron transition at nonzero $T$ and $\mu$ in the 
 presence of magnetic field were considered in the framework of the FCM by the authors of 
Ref.\,\cite{Si7}. 
 In the present work we leave the investigation of strange stars with magnetic field to 
be considered in future works.
 
 We start the work by studying the anomalous mass defects in the first (ascendant) 
branches of the stellar sequences, which also includes the region of low mass strange 
stars.
 To this end, we first use the solutions of the Tolmann-Oppenheimer-Volkov (TOV) 
equations for the case of constant energy density (see Ref.\,\cite{ShT}) to guide our 
investigation. 
 In this case, the ratio pressure-to-energy density, ${\cal P}_C/{\cal E}_C$, 
at the center of a star can be expressed in terms of the surface or gravitational 
redshift (henceforth simply called redshift) of a radiation emitted at a given frequency 
from the star surface.
 Thus, we extend this simple idea to find an analogous description of the general case 
of stars with non-constant energy density profiles. 

 The new description can be used to study strange stars in the low mass region of stellar 
sequences. 
 While masses and radii of strange stars and neutron stars are similar in the high 
mass regions of the mass-radius diagram, the radii of strange stars and neutron stars 
with similar masses are very different in the low mass region (cf. Fig.\,1 in 
Ref.\,\cite{Bom}).  
 On the other hand, masses and radii decrease for large values of the parameters $V_1$ 
and $G_2$  of the FCM nonperturbative equation of state, as hown in Ref.\,\cite{Fla1}.
 So, strange stars of low mass are appropriate for investigations in the region of large 
values of the parameters $V_1$ and $G_2$.
 As an example of the applicability of our description, we take the astrophysical 
observations of the compact star 1E\,1207.4-5204 (which is a low mass one) to determine the 
parameters $V_1$ and $G_2$. 
 A detailed investigation, but within the MIT Bag Model, in Ref.\,\cite{RXX} indicates that 
1E\,1207.4-5204 may be a strange star. 
 In Sec.\,\ref{cmpctst1E1207} we also assume the strange star hypothesis to investigate 
1E\,1207.4-5204 within the FCM. 

 The presentwork is twofold. 
 First, it considers compact stars as laboratories to determine the important quantities 
of the cold SQM at very high densities to describe the low $T$ and high $\mu$ (or density) 
region of the QCD phase diagram. 
 On the other side, there are the experiments at RHIC and LHC to investigate the quark-gluon 
plasma properties in the high temperature region of the phase diagram.
 Second, on the astrophysical side, it serves to investigate strange stars preperties governed 
by a nonperturvative equation of state provided by the FCM. 

 The present paper is organized as follows. 
 In Sec.\,\ref{npeos} we recall the FCM main equations to be used in our calculation.
 In Sec.\,\ref{stc} we present the equations needed to describe the stellar configurations. 
  Sec.\,\ref{nctedpfl} is devoted to the general case of strange stars with non-constant 
internal energy density profile. 
 In Sec.\,\ref{cmpctst1E1207}, we make an attempt to estimate the model parameters $V_1$ 
and $G_2$ from the observations of the compact star 1E\,1207.4-5209\,.  
Sec. \ref{fnlrmrks} is dedicated to the final remarks. 


\section{The nonperturbative equation of state at zero temperature}
\lb{npeos}

 Let us recall the main equations of the FCM thermodynamics of quarks (see 
Refs.\,\cite{Fla1,Fla2,Fla3} for details).
 The main parameters of the nonperturbative equation of state are the large distance 
static $Q{\bar Q}$ potential $V_1$ and the gluon condensate $G_2$\,.
 The pressure, energy density and number density of a quark gas at $T=0$ are given by 
\bqn
p_q&=&\frac{N_c}{3\pi^2}\Bigg\{\frac{k_q^3}{4}\sqrt{k_q^2+m_q^2}-
\frac{3}{8}\;m_q^2\bigg[k_q\sqrt{k_q^2+m_q^2}\nonumber\\
&&-m_q^2\;\ln\bigg(\frac{k_q+\sqrt{k_q^2+m_q^2}}{m_q}\bigg)
\bigg]\Bigg\}\;,
\lb{pqT0}
\eqn
\bqn
\varepsilon_q &=&\frac{N_c}{\pi^2}\Bigg\{\frac{k_q^3}{4}\sqrt{k_q^2+m_q^2}+
\frac{m_q^2}{8}\;\bigg[k_q\sqrt{k_q^2+m_q^2}\nonumber\\
&&-m_q^2\;\ln\bigg(\frac{k_q+\sqrt{k_q^2+m_q^2}}{m_q}
\bigg)\bigg]\nonumber\\
&+&\frac{V_1}{2}\;\frac{k_q^3}{3}
\Bigg\}
\lb{eqT0}
\eqn
and 
\bq
n_q = \frac{N_c}{\pi^2}\;\frac{k_q^3}{3}\;,
\lb{nqT0}
\eq
where
\bq
k_q=\sqrt{(\mu_q-V_1/2)^2-m_q^2}\;,\;\;\;\;(q=\rm{u,d,s})\;,
\lb{kq}
\eq 
$N_c=3$ is the color number. 
 The total pressure, energy density and particle number density including 
electrons are given by
\bq
p=\sum_{q={\rm u,d,s}}p_q-\Delta|\varepsilon_{\rm vac}|+p_{\rm e}\;,
\lb{pqgl}
\eq
\bqn
\varepsilon =\sum_{q={\rm u,d,s}}\varepsilon_q+
\Delta|\varepsilon_{\rm vac}|+\varepsilon_{\rm e}\;,
\lb{eqgl}
\eqn
\bq
n=n_{\rm u}+n_{\rm d}+n_{\rm s} + n_{\rm e}\,, 
\lb{nqg}
\eq
where 
\bq
\Delta|\varepsilon_{\rm vac}|=\frac{11-\frac{2}{3}N_f}{32}\Delta G_2 
\lb{dvac}
\eq
is the vacuum energy density difference between confined and deconfined phases 
(which from now on will be called vacuum energy density), $N_f=3$ is the number 
of flavors, and $\Delta G_2\simeq \frac{1}{2}G_2$ as in Refs.\,\cite{ST1,ST2}.
 The value of the gluon condensate, $G_2=0.012\pm0.006\,{\rm GeV}^4$, has been 
determined by QCD sum rule techniques by the authors of Ref.\,\cite{SVZ}.

To obtain the numerical equivalence for the MIT Bag Model we take 
$\Delta|\varepsilon_{\rm vac}|=B$ and $V_1=0$\,. 
 However, we emphasize that $\Delta|\varepsilon_{vac}|$ is a nonperturbative quantity. 
 In the present work, we assume that $V_1$ and $G_2$ are constant quantities (i. e., 
independent on any flavor chemical potential, or density).
 We also assume that the SQM composition of strange stars satisfies the requirements 
of chemical equilibrium under the weak interactions and charge neutrality to 
perform our calculations of the stellar configurations, as in Refs.:\,\cite{Fla1,Fla2,Fla3}.

 We use the following values for the quark masses: $m_u=5$ MeV, $m_d=7$ MeV and 
$m_s=150$ MeV. 
  The equations for the degenerate electron gas contribution are obtained 
by making the changes: $N_c\rightarrow1$, $V_1\rightarrow0$, $\mu_q\rightarrow\mu_\ex$ 
and $m_q\rightarrow m_\ex$\,, in Eqs.\,(\ref{pqT0})-(\ref{nqT0}). 
 
\section{stellar configurations} 
\lb{stc}

 Stellar configurations are calculated by numerical integration of the 
hydrostatic equilibrium equations of Tolman-Oppenheimer-Volkov as in Refs.\,\cite{ShT,Gle} 
(we come back to these equations in Sec.\,\ref{lmsstrs}).
 Of special importance is the total gravitational mass of a compact star 
(see Ref.\,\cite{ZeN} for details),  
\bq
M=\int^R_0\varepsilon(r)\,dv(r)\;,
\lb{stm}
\eq
where $dv(r)=4\pi r^2dr$, which is the mass that governs the Keplerian orbital 
motion of the distant gravitating bodies around it, as measured by external 
observers. 
 The baryonic mass (also called rest mass) of a star is $M_A=m_A\,N_A$, where 
$m_A$ is the mass per baryon of the baryon specie $A$, with the number of baryons 
given by
\bq
N_A=\int^R_0n_A(r)\,\ex^{\lambda(r)}\,dv(r)\,,
\lb{NA}
\eq 
where $\ex^{\lambda(r)}=[1-2\,GM(r)/(c^2r)]^{-1/2}$  is the spatial function of the 
metric; $M(r)$ is the mass within a sphere of radius $r$, and 
\bq
n_A=\frac{1}{3}(n_{\rm u}+n_{\rm d}+n_{\rm s}) 
\lb{nA}
\eq
is the equivalent baryon number density.
 The baryonic mass is the mass that the star would have if its baryon content were 
dispersed at infinity, with zero kinetic energy.
 In the case of strange stars (because of the quark confinement), $N_A$ is the 
equivalent number of baryons (not quarks). 
 We here assume $m_A=m_n$ as in our previous works in Refs.:\,\cite{Fla1,Fla2,Fla3}.  


\subsection{Mass defects}
\lb{mdfcts}

 With the masses $M$ and $M_A$ known, we are in a position to calculate the mass defect  
(which in our notation\footnote{We here follow the notation according to 
Refs.\,\cite{Amb1,Amb2,Sak,Var1,Var3}} is minus the binding energy $E_b$ defined in 
Refs.\,\cite{ZeN,Gle}) given by 
 \bq
\Delta_2M=M_A-M\,.
\lb{d2M}
\eq 
 The mass defect corresponds to the energy released to aggregate from infinity the 
dispersed baryonic matter.
 In the present work, we will consider unstable stellar configuration with 
$\Delta_2M<0$ (anomalous mass defect)\,. 
 In this case, the stellar  configuration has an energy excess with respect to 
the energy it would have to be a bound system. 
 The star is in a metastable state, it might explode or implode in the presence 
of an external perturbation.

 In order to consider the general aspects of the anomalous mass defects, let us 
write the differential elements of the total gravitational mass, the equivalent 
number of baryons, and the baryonic (or rest) mass given by  
\bq
dM(r)=\varepsilon\,dv(r)
\lb{dM}
\eq
\bq
dN_A(r)=n_A(r)\,\ex^{\lambda(r)}\,dv(r)
\lb{dNA}
\eq
\bq
dM_A(r)=m_A\,dN_A(r)\,.
\lb{dMA}
\eq
 Then we find that
\bqn
\frac{dM(r)}{dN_A(r)} &=& \frac{\varepsilon(r)}{n_A(r)}\,\ex^{-\lambda(r)} \nonumber\\
&=& m(r)\sqrt{1-2\frac{G}{c^2}\frac{M(r)}{r}}\,.
\lb{dMNAr}
\eqn
We call $m(r)\equiv\varepsilon(r)/n_A(r)$ the mass-energy per baryon inside the star. 
 On the surface of the star (taking into account that $\ex^{-\lambda_R}=\ex^{\phi_R}$ 
at $r=R$)\, Eq.\,(\ref{dMNAr}) becomes
\bqn
\frac{dM(r)}{dN_A(r)}\bigg|_{r=R} &=& m_R\,\ex^{\phi_R} \nonumber\\
&=& m_R\sqrt{1-2\frac{G}{c^2}\frac{M}{R}}\,,
\lb{dMNAR}
\eqn
where $\phi$ is the temporal function of the metric, $m_R\equiv m(R)$ and $M\equiv M(R)$ 
given by Eq.\,(\ref{stm}). 
 
\begin{figure*}[th]
	\centering
	{\includegraphics[scale=0.4]{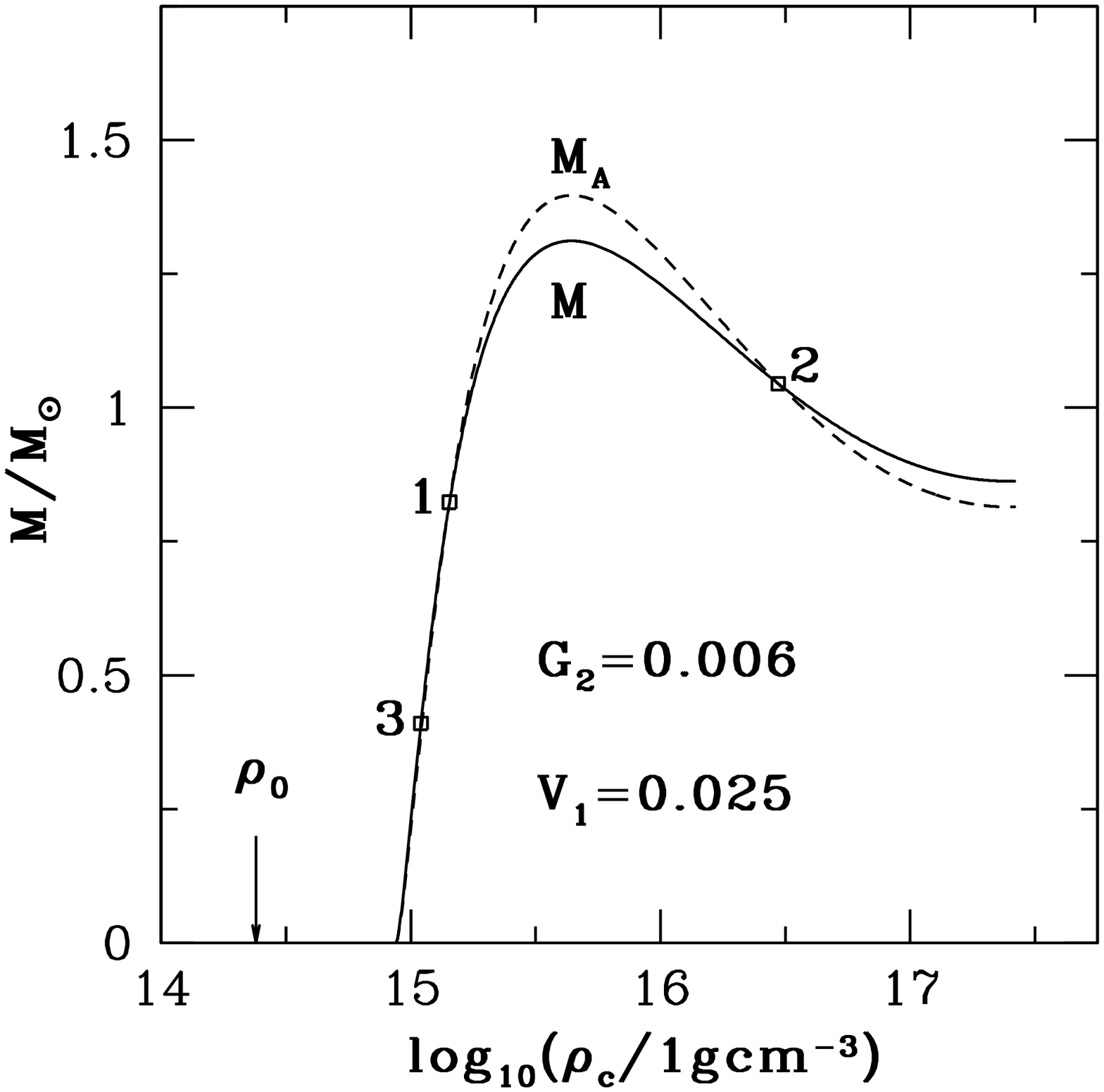}}
\caption{ 
 For the given values of $G_2$ (in ${\rm GeV}^4$ units) and $V_1$ 
(in ${\rm GeV}$ units), gravitational mass $M$ and baryonic mass 
$M_A$ (in units of the solar mass) as functions of the central density. 
 The arrow indicates the nuclear energy density 
 $\rho_0\simeq2.5\times10^{14}\,{\rm g\,cm^{-3}}$\,. 
}
\label{mNd2m}
\end{figure*}

\begin{figure*}[th]
	\centering
	{\includegraphics[scale=0.4]{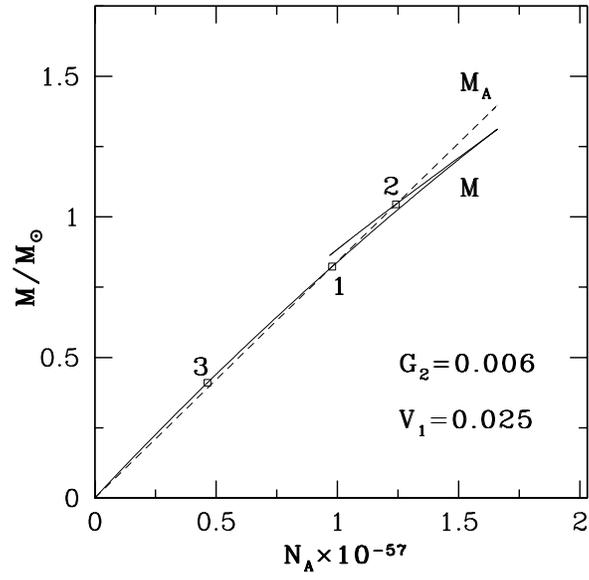}}
\caption{ 
 Same as the previous figure but for gravitational mass $M$ and baryonic mass $M_A$ 
versus baryonic number $N_A$\,.
  Points {\bf 1}, {\bf 2} and {\bf 3} are in direct correspondence with  
the ones in Fig.\,\ref{mNd2m}. 
}
\label{mNd2ma}
\end{figure*}

\begin{figure*}[th]
	\centering
	{\includegraphics[scale=0.4]{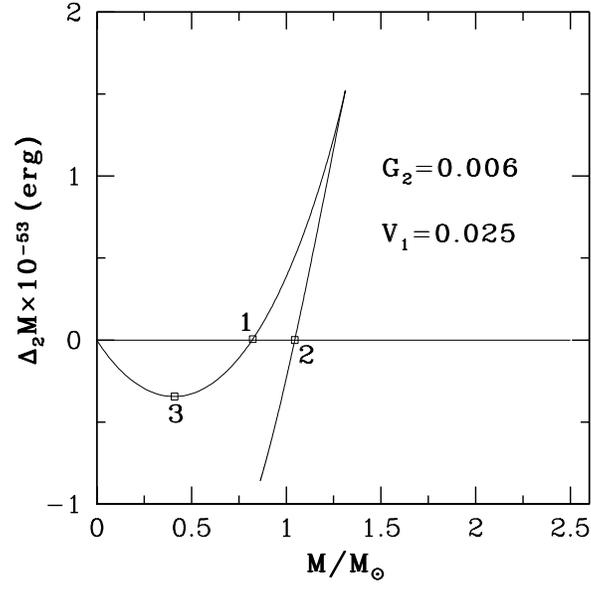}}
\caption{ 
Same as the previous figures but for mass defect versus gravitational mass $M$. 
  Points {\bf 1}, {\bf 2} and {\bf 3} are in direct correspondence with  
the ones in Figs.\,\ref{mNd2m} and \ref{mNd2ma}. 
}
\label{mNd2m2}
\end{figure*}

 Fig.\,\ref{mNd2m} shows the general features of the stellar sequence for the 
given values of the parameters $V_1$ and $G_2$. 
 The point {\bf 3} (in the first branch) indicates an intermediate point around 
$0.4M_\odot$ in the low mass region, where the mass defect is anomalous 
(more visible in Figs.\,\ref{mNd2ma} and \ref{mNd2m2}). 
 In Fig.\,\ref{mNd2ma}, the slope of the $M$ vs. $N_A$ curve at the point {\bf 3} given 
by Eq.\,(\ref{dMNAR}) and that of the $M_A$ vs. $N_A$ plot are parallel, which 
means that 
\bqn
m_R &=& m_A\,\ex^{-\phi_R} \nonumber\\
    &=& m_A(1+z_S)\,,
\lb{MRMAZS}
\eqn
which is a redshift relation connecting the mass per baryon at the surface of 
the star with the baryonic mass $m_A$ taken as a reference mass by a distant 
observer. 
 Once $m_A$ is given, we obtain $m_R$ by measuring $z_S$\,. 
 On the other hand, inside the star, the energy density $\varepsilon(r)$, the 
baryonic number density $n_A(r)$, and the baryonic mass per baryon $m(r)$ decrease 
from the center to the surface of the star. 
 Thus, the inequalities $m_C>m_R>m_A$ (where $m_C\equiv m(0)$) hold in the star 
interior. 
 In other words, there is anomalous mass defect when $m(r)>m_A$\,\,$\forall\,\,r\in[0,R]$. 
 If $m_R\rightarrow m_A$ then the point {\bf 3} goes to the origin of the $M$ vs. $N_A$ 
plot in Fig.\,\ref{mNd2ma} indicating the absence of the anomalous mass defect when 
$m_R=m_A$, in the first branch of the stellar sequence, as shown in 
Refs.\,\cite{Amb1,Amb2,Amb3,Sak,ZeN}. 
 Moreover, for a given equation of state characterized by fixed values of $V_1$ 
and $G_2$\,, $m_R$ is constant for all stars along the corresponding stellar sequence. 
 The slope $m_R\,\ex^{\phi_R}$ in Eq.\,(\ref{dMNAR}) evolves according to the temporal 
function of the metric  along the stellar 
sequence, being greater than $m_A$ at the origin of the sequence and less than $m_A$ 
everywhere above the point {\bf 3}.
 Fig.\,\ref{mNd2m2} shows the mass defect. 
 We see by simple inspection that $|\Delta_2M|$ is maximum at the point {\bf 3}. 

\begin{figure}[th]
	\centering
	{\includegraphics[scale=0.4]{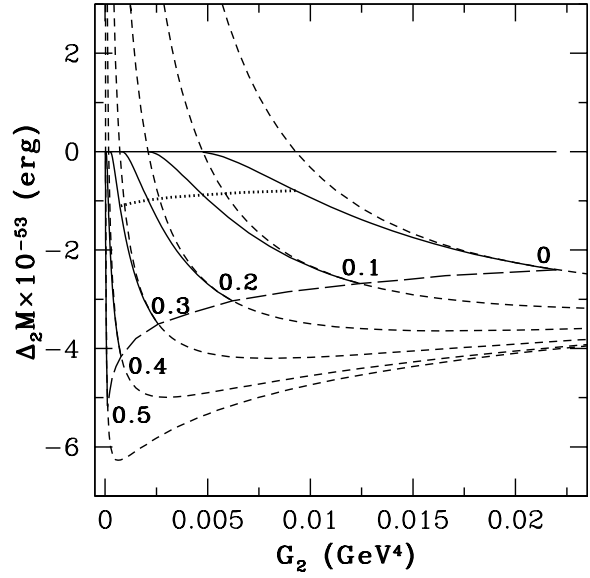}}
	\caption{Mass defect $\Delta_2M$ versus $G_2$ for different values of $V_1$ 
between $V_1=0$ and $V_1=0.5$\,GeV. 
 The labels correspond to the values of $V_1$ (in GeV units). 
 {\it Solid lines}: $\Delta_2M$ at the point {\bf 3} with $m_R$ and $m_A$ satisfying 
Eq.\,(\ref{MRMAZS}).
 {\it Short dashed lines}: $\Delta_2M$ at the maximum masses of the stellar configurations 
(cf. Fig. 4 in Ref.\,\cite{Fla3}).
 {\it Dotted line}: $\Delta_2M$ at the point {\bf 3} of the low stellar masses ranging 
from $M/M_\odot\simeq0.6$ at $V_1=0$\, to $M/M_\odot\simeq1.1$ at $V_1=0.3$\,GeV\,.
 {\it Long dashed line}: only connecting the points of $\Delta_2M$\, where the point {\bf 3} 
is at the maximum mass.
 }
\label{ponto3}
\end{figure}

 We have calculated for different values of $V_1$ and $G_2$ the maximum values of 
$|\Delta_2M|$ at the point {\bf 3} by searching for points where Eq.\,(\ref{MRMAZS}) 
holds. 
 The results are depicted in Fig.\,\ref{ponto3}. 
 Each plot starts from the origin ($\Delta_2M=0$) and ends at the maximum mass 
value of $\Delta_2M$ calculated in Ref.\,\cite{Fla3} where the point {\bf 3} is located 
(also satisfying Eq.\,(\ref{MRMAZS})).
 In the low mass region (connected by the dotted line) $|\Delta_2M|$ is around 
$1\times10^{53}$\,erg. 
 For the sake of comparison, the energies liberated in type Ia Supernovae originated 
by white dwarf explosions are of the order of $(1-2)\times10^{51}$\,erg\,.
 The total energy of the observed neutrinos in the supernova 1987A was found to be around 
$\sim3\times10^{53}$\,erg\,.
 Thus, explosions of strange stars with anomalous mass defects could be a possibility to 
be considered.
 
\subsection{The Tolmann-Oppenheimer-Volkov equations}
\lb{lmsstrs}

 To simplify our notation let us define the dimensionless radius 
and mass by
\bq
{\cal X}\,\equiv\,\frac{c^2}{GM_\odot}\,r\,
~~~~{\rm and}~~~~{\cal Z}\,\equiv\,\frac{M(r)}{M_\odot}\,, 
\lb{calXZ}
\eq
where $M(r)$ is the mass within the sphere of radius $r$; and 
$GM_\odot/\,c^2\simeq1.5$\,km.
 Thus, the TOV equations are given by  
\bqn
\frac{d\,{\cal Z}}{d\,{\cal X}}&=&\eta\,{\cal X}^2\,{\cal E}\,,
\lb{tov4}
\eqn
\bqn
\frac{d\,{\cal P}}{d\,{\cal X}}&=&-\,\frac{(\,{\cal E}+{\cal P}\,)\,
(\,{\cal Z}+\eta\,{\cal X}^3\,{\cal P}\,)}
{{\cal X}^2\,(\,1-2\,\,{\cal Z}\,/\,{\cal X}\,)}\,,
\lb{tov5}
\eqn
where 
$\eta\equiv\,4\pi\,(GM_\odot/c^2)^3/M_\odot c^2\simeq0.03628$ ${\rm fm^3\,GeV^{-1}}$.
 The redshift of the spectral lines emitted from the star surface is given by
\bq
z_S = \frac{1}{\sqrt{1\,-\,2\,{\cal Z}_R/{\cal X}_R}}-1\,,
\lb{zs1}
\eq
where ${\cal X}_R\equiv{\cal X}(r=R)$ and ${\cal Z}_R\equiv{\cal Z}({\cal X}_R)$. 

   As the redshift has an important role in the present work, let us 
explore some properties which will be important in the sequel.
 For finite values of $\varepsilon_C$ and $p_C$ at the center of the star\footnote{
Where ${\cal Z}({\cal X})/{\cal X}\rightarrow0$ when
${\cal X}\rightarrow0$\,.} we find  
\bq 
\frac{d\,{\cal P}({\cal X})}{d\,{\cal X}}\bigg|_{{\cal X}=0}=0\,.  
\lb{dpdfi0}
\eq
 At the star surface where energy the density is ${\cal E}_R$ and ${\cal P}_R=0$ 
we have
\bqn
\frac{1}{{\cal E}_R}\frac{d\,{\cal P}({\cal X})}{d\,{\cal X}}\bigg|_{{\cal X}={\cal X}_R}
  &=&-\,\frac{{\cal Z}_R\,/{\cal X}_R }{{\cal X}_R\,[~1-2\,{\cal Z}_R\,/\,{\cal X}_R~]}\,\nonumber\\
  &=&-\,\frac{z_S^2 + 2\,z_S } {2\,{\cal X}_R}\,\nonumber\\
  &=&-\,\frac{(z_S^2 + 2\,z_S)^2 } {4\,{\cal Z}_R\,(1+z_S)^2}\,.
\lb{dpR}
\eqn
 These expressions are of general validity for constant and non-constant energy densities.
 The right hand sides of Eqs.\,(\ref{dpR}) are observables quantities directly given in terms 
of $z_S$ and the dimensionless radius ${\cal X}_R$ or mass ${\cal Z}_R$\,.
 
\subsection{Constant energy density}
\lb{ctnrgdnst}

 In the case ${\cal E}\equiv{\cal E}_C=$\,cte. (see Ref.\,\cite{ShT}) we find that
\bq
{\cal Z}\,=\,\frac{1}{3}\,\eta\,{\cal E}_C\,{\cal X}^3\,,
\lb{Zmass}
\eq
\bqn
{\cal P}\,&=&\,{\cal E}_C\frac{\sqrt{1-2{{\cal Z}}/{\cal X}}-\sqrt{1-2{{\cal Z_R}}/{\cal X_R}}}
{3\,\sqrt{1-2{{\cal Z_R}}/{\cal X_R}}-\sqrt{1-2{{\cal Z}}/{\cal X}}}\nonumber\\
  &=&\,{\cal E}_C\frac{\sqrt{1-\frac{2}{3}\,\eta\,{\cal E}_C\,{\cal Z}\,{\cal X}^2}-
                     \sqrt{1-\frac{2}{3}\,\eta\,{\cal E}_C\,{\cal Z}_R\,{\cal X}_R^2}}
{3\,\sqrt{1-\frac{2}{3}\,\eta\,{\cal E}_C\,{\cal Z}_R\,{\cal X}_R^2}-
    \sqrt{1-\frac{2}{3}\,\eta\,{\cal E}_C\,{\cal Z}\,{\cal X}^2}}\,.
\lb{Pc}
\eqn
 The redshift is now given by
\bq
z_S = \frac{1}{\sqrt{1\,-\frac{2}{3}\,\eta\,{\cal E}_C\,{\cal Z}_R\,{\cal X}_R^2}}-1\,.
\lb{zs2}
\eq

 Taking ${\cal X}=0$ in Eqs.\,(\ref{Pc}), the ratio ${\cal P}_C/{\cal E}_C$ 
at the center of the star is given by   
\bq
\frac{{\cal P}_C}{{\cal E}_C}\,=\,\frac{z_S}{2-z_S}\,. 
\lb{PezS}
\eq
The equation of state at the center of a compact star with constant energy density 
can be obtained by direct redshift measurements.
 Thus, we can use  the redshift as a probe to give us the equation of state  
at the center of a compact star in the ${\cal E}\equiv{\cal E}_C=$cte. 
approximation.
 For finite ${\cal P}_C/{\cal E}_C\geq0$, we note that $z_S<2$, according 
to Eq. (11.6.20) in Ref.\,\cite{StW}. 
  Additionally, we note that the above solution for ${\cal P}_C/{\cal E}_C$ does 
not depend on the equation of state of the nuclear or strange matter. 
 This is an interesting property to be used to test theoretical models. 

Low mass compact stars have been commonly accepted as the ones with masses lower 
than the solar mass, and characterized by almost constant internal energy density 
profiles. 
 The masses of low mass strange stars can be calculated by the Newtonian approximation 
$M\simeq\,(4\,\pi/3)\,\varepsilon_S\,R^3$\,, where $\varepsilon_S$ is the surface 
energy density, as in Ref.\,\cite{HPY}. 
 However, not all low mass compact stars can be approximated by constant internal 
energy density profiles, as we show in Sec.\,\ref{cmpctst1E1207}. 
 For instance, for certain values of the FCM parameters $V_1$ and $G_2$, the shape 
of the energy density may present a remarkable change from the center to the surface 
of the star. 
 In this case, we do not have an analogous prescription to the one given in 
Eq.\,(\ref{PezS}). 
 However, we can explore the behavior of the theoretical dependence of 
$({\cal P}/{\cal E})_C$ on $z_S$ to find a corresponding expression for the case of 
non-constant energy density profile. 



\section{\bf $({\cal P}/{\cal E})_C$ for non-constant energy density}
\label{nctedpfl}

\begin{figure}[th]
	\centering
	{\includegraphics[scale=0.4]{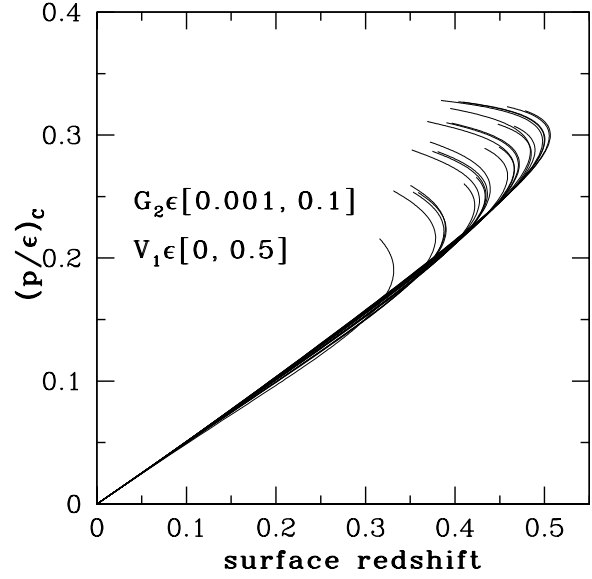}}
	\caption{ 
 For the given ranges of $G_2$ (in GeV$^4$ units) and  $V_1$ (in GeV units), 
ratios $({\cal P}/{\cal E})_C$ as function of the redshift. 
 The values of $V_1$ increase from top to bottom. 
 The values of $G_2$ increase from bottom to top.
}
\label{pipifig}
\end{figure}

\begin{figure}[th]
	\centering
	{\includegraphics[scale=0.4]{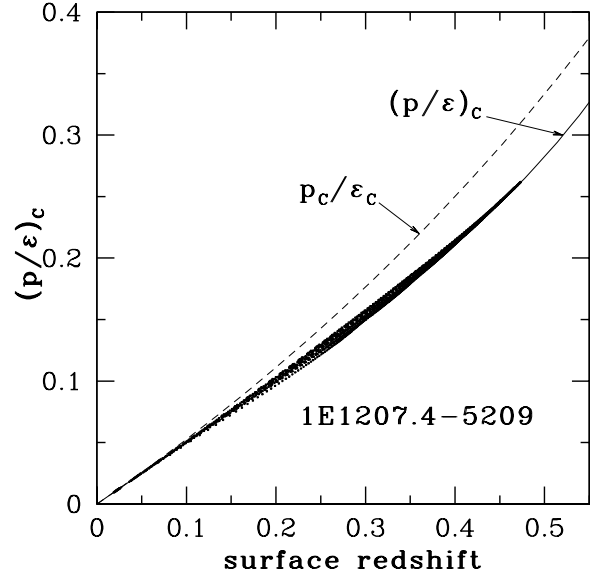}}
	\caption{ 
 Same as the previous figure but for the values of $({\cal P}/{\cal E})_C$ 
ending at the maximum masses of the stellar configurations.
 {\it Short dashed line}\,: ${\cal P}_C/{\cal E}_C$ given by Eq.\,(\ref{PezS})\,.
 {\it Solid line}\,: $({\cal P}/{\cal E})_C$ given by Eq.\,(\ref{pipi2}) 
extrapolated for higher values of the redshift to become visible.
}
\label{pipifiga}
\end{figure}

 Let us now try to build a representation for  $({\cal P}/{\cal E})_C$ to 
simulate a general case when the energy density is not constant. 
 To this end, we first generate many (theoretical) stellar sequences, each one 
for a fixed pair of parameters ($V_1$\,,\,$G_2$). 
 The plots of the corresponding ratios $({\cal P}/{\cal E})_C$ vs. $z_S$ are shown 
in Fig.\,\ref{pipifig}.
 We observe that all curves that start very close together from the origin, in a thin 
bundle of lines, deviate from the initial direction at certain points along the 
bundle  resembling a "cockatiel crest" at the upper parts of the figure. 
 Moreover, the deviation points, each one corresponding to a pair ($V_1\,,\,G_2$), are 
located at the maximum masses of the corresponding stellar sequences. 
 The lines of the "crest" in the second branches of the stellar sequences 
(such as the one with the point {\bf 2}\,, in Fig.\,\ref{mNd2m}) are of no interest 
in the present work.
 So, removing the lines of the "crests" at the points of maximum masses, we obtain 
a thin cloud of aligned points which converge in the low redshift region 
(say, $z_S\gappl0.1$) to the constant energy density solution given by Eq.\,(\ref{PezS})\,, 
as shown in Fig.\,\ref{pipifiga}\,.  
 
\begin{figure}[th]
	\centering
	{\includegraphics[scale=0.4]{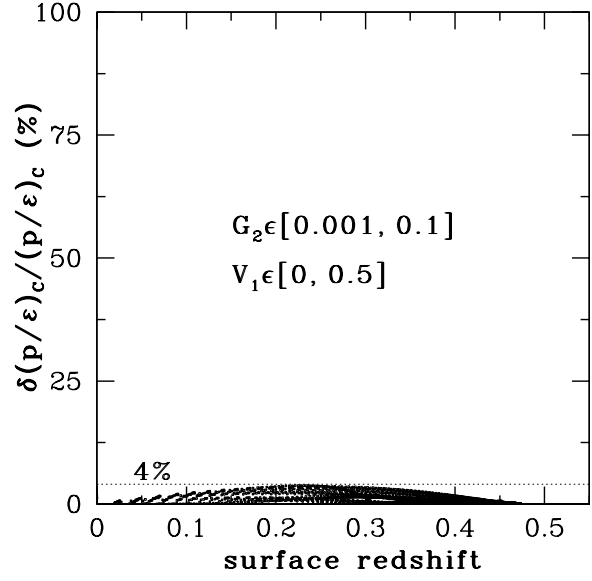}}
	\caption{ 
 For the given ranges of $G_2$ (in GeV$^4$ units) and  $V_1$ (in GeV units), 
fractional error $\delta({\cal P}/{\cal E})_C\,/\,({\cal P}/{\cal E})_C$ 
as function of the redshift.
}
\label{pipifigb}
\end{figure}

\begin{figure}[th]
	\centering
	{\includegraphics[scale=0.4]{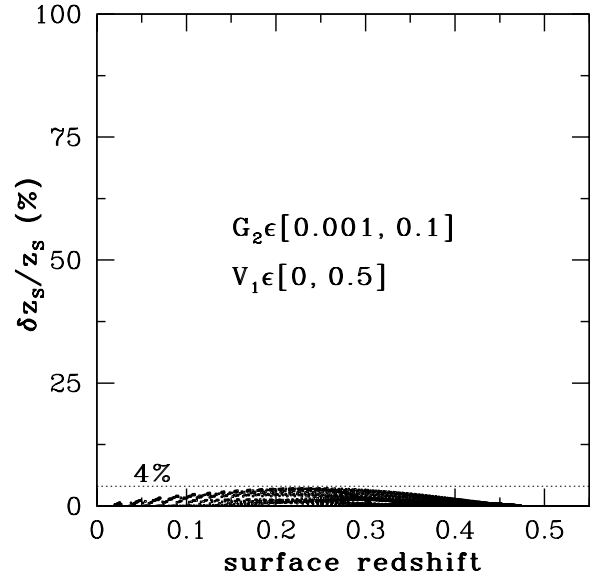}}
	\caption{ 
 Same as Fig.\,\ref{pipifigb} but for $\delta z_S/z_S$\,.
}
\label{pipifigc}
\end{figure}

 The next step is to represent the cloud of points by the interpolating curve   
given by 
\bq
\bigg(\frac{{\cal P}}{{\cal E}}\bigg)_C\,=y\,(1-\frac{9}{8}\,y + 2\,y^2) 
\lb{pipi2}
\eq
where $y\equiv{\cal P}_C/{\cal E}_C$ is the constant energy density solution given 
by Eq.\,(\ref{PezS}).
 In Fig.\,\ref{pipifiga}\,, the solid line shows the interpolating curve extrapolated 
to higher redshifts to become visible. 
 The second and third coefficients on the right hand side of Eq.\,(\ref{pipi2}) were 
initially determined by best-fitting methods (see Refs.:\,\cite{Bev1,Bev2}) and then 
rounded in order to give an error estimate $\gappl4$\% (the best we obtained after 
many attempts!) at an intermediate redshift range, and zero errors at zero redshift 
and at the maximum mass redshift, as shown in Figs.\,\ref{pipifigb} and \,\ref{pipifigc}. 
 The fractional error of $({\cal P}/{\cal E})_C$ as function of 
the redshift and the corresponding error for $z_S$ in terms of $({\cal P}/{\cal E})_C$ 
(obtained by inverting Eq.\,(\ref{pipi2}))\,, respectively, are shown in a scale from 
zero to 100.
 Thus, by this way, we are able to use a redshift measurement as a probe to estimate 
$({\cal P}/{\cal E})_C$ at the center of a strange star within the errors shown in 
Figs.\,\ref{pipifigb} and \,\ref{pipifigc}.
 Of course, this is a model dependent procedure valid for the case of the FCM 
nonperturbative equation of state we are considering, but with an interesting 
quasi-model independent feature.
 An application of Eq.\,(\ref{pipi2}) is done in Sec.\,\ref{cmpctst1E1207} to 
investigate the compact star 1E\,1207.4-5209.
 
\begin{figure}[tbh]
\centering
	{\includegraphics[scale=0.4]{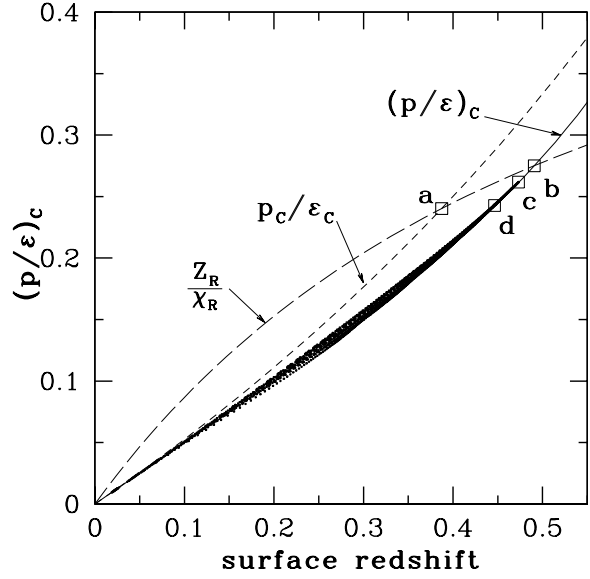}}
	\caption{
 Same as Fig.\,\ref{pipifigb} but including the ${\cal Z}_R/{\cal X}_R$ plot 
 (long dashed line). 
 The point {\bf a} corresponds to Eq.\,(\ref{pcec1}) and the point {\bf b} corresponds 
to Eq.\,(\ref{pcec2}).
 For $V_1=0$, the point {\bf c} indicates the upper bound $({\cal P}/{\cal E})_{C,\rm\bf c}\simeq0.262$ and the corresponding redshift $z_S\simeq0.47$\,.
 For $V_1=0.5$ GeV, the point {\bf d} indicates the upper bound 
 $({\cal P}/{\cal E})_{C,\rm\bf d}\simeq0.24$ and the corresponding redshift 
 $z_S\simeq0.44$\,.
 }
\label{pcecZX}
\end{figure}

\begin{figure}[th]
\centering
	{\includegraphics[scale=0.4]{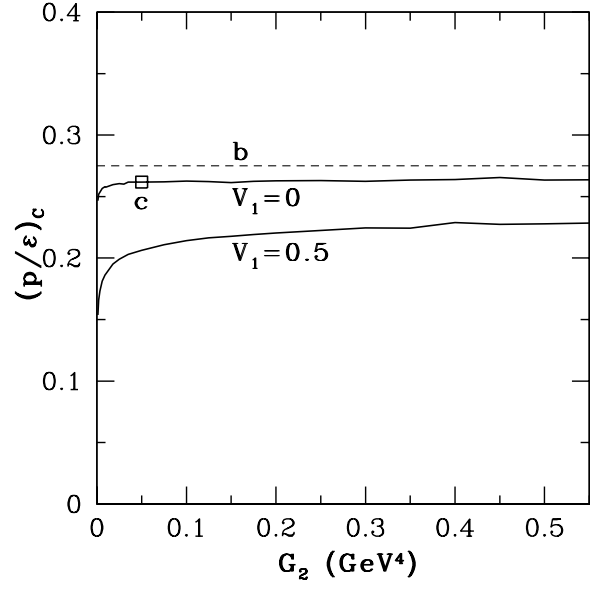}}
	\caption{
 For increasing values of $G_2$, the "constancies" of $({\cal P}/{\cal E})_C$ 
at {\bf b} and {\bf c}. 
 For $V_1=0.5$ GeV, the "constancy" at the point {\bf d} is not visible in the 
scale of the figure.
}
\label{pcecZXa}
\end{figure}

 Coming back to Fig.\,\ref{pipifig}, the curves at the upper parts of the figure become 
more and more closer, but never exceed the limiting redshift $z_S\simeq0.51$\,.   
 On the other hand, depending on the values of $V_1$, the values of 
$({\cal P}/{\cal E})_C$ at the maximum masses along the bundle of curves do  
not exceed a certain limit whatever the values of $G_2$ may be, suggesting the 
existence of an upper limit for $({\cal P}/{\cal E})_C$ and a corresponding limit 
for $G_2$. 
 In order to obtain the limits for $G_2$, we have considered the solutions for 
the cases with $V_1=0$ (which gives the highest "crests"), and $V_1=0.5$\,GeV 
(which gives the lowest "crests"). 
 In Fig.\,\ref{pcecZX}, for $V_1=0$ and $G_2\gappr0.035 \,{\rm GeV}^4$,  
$({\cal P}/{\cal E})_{C,\,\rm\bf c}$ (at the point {\bf c}) becomes constant around 
$0.262$ at $z_S\simeq0.47$ as it is indicated by the point {\bf c} in Fig.\,\ref{pcecZXa}. 
 At this point we have ${\cal Z}_R/{\cal X}_R\equiv R_S/(2\,R)\simeq0.27$, 
where $R_S=2\,GM/c^2$ is the Schwarzschild radius of the star. 
 The mass of the star is $M\simeq0.58\,M_\odot$ and its radius is 
$R\simeq3.19\,{\rm km}\simeq1.85\,R_S$\,.
 For $V_1=0.5$\,GeV,  $({\cal P}/{\cal E})_{C,\,\rm\bf d}$ saturates 
around $0.24$ at $z_S\simeq0.44$, but at a too large value of $G_2$, say, 
$\gappr2.5\,{\rm GeV}^4$ (the point {\bf d} is not shown in Fig.\,\ref{pcecZXa})\,.  
 The corresponding mass and radius are  $M\simeq0.06\,M_\odot$ and 
$R\simeq0.34\,{\rm km}\simeq1.92\,R_S$.

  Let us now consider the the particular situation given by  
\bqn
\bigg(\frac{{\cal P}}{{\cal E}}\bigg)_C&=&\,\frac{{\cal Z}_R}{{\cal X}_R}\nonumber\\
&=&\frac{1}{2}\bigg[1-\frac{1}{(1+z_S)^2}\bigg]\,.
\lb{pexz}
\eqn
 Fig.\,\ref{pcecZX} also shows the plot of the right hand side of Eq.\,(\ref{pexz}) 
together with the plots of ${\cal P}_C/{\cal E}_C$ and  $({\cal P}/{\cal E})_C$. 
 By the logic of Fig.\,\ref{pipifig} the points of the plot given by Eq.\,(\ref{pexz}) 
are located on the second branches of the stellar sequences.
  The point {\bf a} is determined by solving the equation 
\bq
y-\frac{1}{2}\bigg[1-\frac{1}{(1+z_S)^2}\bigg]=0\,, 
\lb{raiz1}
\eq
from which we obtain the root $z_S\simeq0.39$ and the corresponding ratio
\bq
\bigg(\frac{{\cal P}}{{\cal E}}\bigg)_{C,\,\rm\bf a}\simeq0.24\,.
\lb{pcec1}
\eq
  Although the point {\bf a} is on the ${\cal P}_C/{\cal E}_C$ plot, the value of 
$({\cal P}/{\cal E})_{C,\,\rm\bf a}$ can also be obtained for certain values of $V_1$ 
and $G_2$. 
 As a result, the point {\bf a} is on the second branches of the stellar sequences 
for values of the pair ($V_1\,,\,G_2$) between 
$(V_1\simeq0.23\,{\rm GeV}\,,\,G_2=0.001\,{\rm GeV}^4)$ and 
$(V_1=0.5\,{\rm GeV}\,,\,G_2\simeq0.012\,{\rm GeV}^4)$ 
 (see the value of the gluon condensate in Ref.\,\cite{SVZ}).
 In the case of the FCM the internal energy densities of the stars within these values  
of ($V_1\,,\,G_2$) are not necessarily constant along the respective stellar sequences. 
 However, roughly constant energy densities occur for masses and radii very lower than the 
ones at the maximum mass. 

 Analogously, by solving the equation
\bq
y(1-\frac{9}{8}y+2y^2)-\frac{1}{2}\bigg[1-\frac{1}{(1+z_S)^2}\bigg]=0\,,
\lb{raiz2}
\eq
we find the root $z_S\simeq0.49$\, and the ratio
\bq
\bigg(\frac{{\cal P}}{{\cal E}}\bigg)_{C,\,\rm\bf b}\simeq0.275\,
\lb{pcec2}
\eq
corresponding to the point {\bf b}. 
 This point is located on the second branches of the stellar sequences for 
$V_1=0$ and $\forall\,\,G_2\gappr0.035\,{\rm GeV}^4$\,. 
 The differences between the values of $({\cal P}/{\cal E})_C$ and $z_S$ at the points 
{\bf b} and {\bf c} are about (4-5)\,\%\,; and the differences for the masses, radii 
and ${\cal Z}_R/{\cal X}_R$ are $\gappl$(1-3)\,\%\,. 
 Thus, for observations with error bars $\gappr$(4-5)\,\% we can assume that 
{\bf b}\,$\simeq$\,{\bf c} ({\bf c} is at the maximum mass) to check, from $z_S$ 
measurements, if the ratio $({\cal P}/{\cal E})_{C,\,\rm\bf c}$\, of an observed strange 
star candidate is near its maximum value.


\section{The compact star 1E\,1207.4-5209}
\label{cmpctst1E1207}

 As mentioned earlier, the low mass strange stars are more appropriate to test the 
applicability of the above theoretical developments. 
 Due to the confinement mechanisms controlled by the parameters $V_1$ and $G_2$, 
the masses and radii of strange stars could be very small (say, $M\gappl0.5\,M_\odot$ 
and $R\gappl10$\,km).
 We have shown in Fig.\,8 of the Ref.\,\cite{Fla1} that the masses and radii of strange 
stars decrease with the increase of $V_1$ and/or $G_2$\,.   
 For instance, the decrease of the maximum mass is pronounced for $V_1=0.5\,{\rm GeV}$ 
and $G_2=0.007\,{\rm GeV}^4$ as shown in Fig.\,2 of the Ref.\,\cite{Fla3}\,. 
 Thus, large values of the model parameters are compatible with low masses and radii.
  On the other hand, we must have in mind that not all compact stars are low mass stars 
and have their masses, radii and redshifts given by the observational data. 
 All this facts make the compact star 1E\,1207.4-5209\, a good example to test the 
above method. 

 The compact star 1E1207.4-5209 is an isolated neutron star (INS) discovered 
near the center of the supernova remnant (SNR) PKS 1209-51 (also known as G296.5+10.0) 
by the Einstein observatory, in Ref.\,\cite{HeB}. 
 Its age is estimated from the remnant, by the authors of Ref.\,\cite{Rog}, to be around 7 kys , with an uncertainty of a factor 3.
 The distance to the remnant is about d=1.3 - 3.9 kpc, as estimated in Ref.\,\cite{Gia}. 
 
 The possible SQM composition of 1E\,1207.4-5209 was considered in Ref.\,\cite{RXX}.  
 In a mass-radius relation investigation in Ref.\,\cite{Zha}, the mass, radius 
and redshift of 1E\,1207.4-5209 were determined by independent methods to be  
$M=0.34\pm0.09\,M_\odot$\,, $R=4.2\pm0.1$\,km and $z_S=0.12-0.23$. 
  Assuming the SQM composition of 1E\,1207.4-5209 governed by the FCM nonperturbative 
equation of state, we now try to find the parameters $V_1$ and $G_2$ from the given 
astrophysical data. 
 Because the mass, radius and redshift are related by Eq.\,(\ref{zs1}) we have two 
independent quantities to determine the parameters.
 As the redshift measurement is given only within a range, we try to obtain the parameters 
by following two steps.

\begin{figure}[th]
	\centering
        {\includegraphics[scale=0.4]{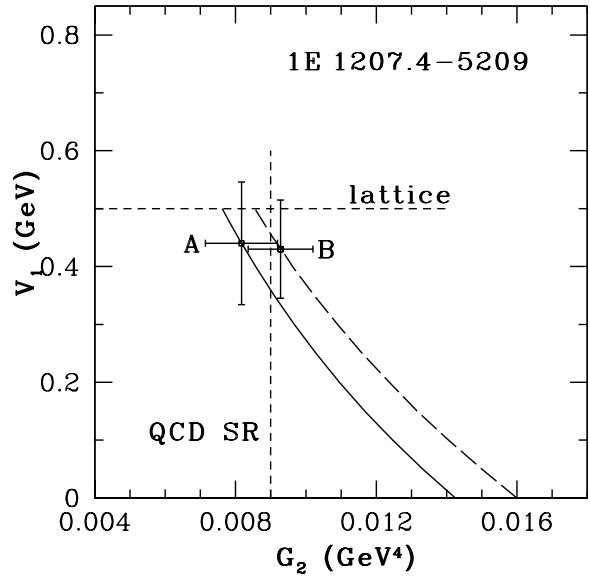}}
	\caption{ 
 Solid and long dashed lines show the first and second attempts to determine 
the model parameters $V_1$ and $G_2$ from the mass, radius and redshift measurements 
provided by the observations of the compact star 1E\,1207.4-5209. 
 The crosses {\bf A} and {\bf B} indicate the final results of our narrowed searches for 
$V_1$ and $G_2$\,. 
 For comparison, the horizontal short dashed line at $V_1=0.5$\,GeV displays the result 
obtained by lattice calculations, in Ref.\,\cite{KaZ}\,. 
 The vertical short dashed line at $G_2=0.009\,{\rm GeV^4}$ indicates the central value of 
$G_2$ obtained by QCD sum rules (QCD SR) in Ref.\;\cite{Iof}.
}
\label{1E1207a}
\end{figure}

 First, we take as the central value of the redshift the one calculated from the values 
of the above mass and radius plus their respective error bars, using 
Eq.\,(\ref{zs1}). 
 Thus, we here assume $z_S=0.15^{+0.057}_{-0.048}$ to be used in our calculation.  
 However, the measurements of $M$ and $R$, which appear in the ratio $M/R$, are not 
suficient to discriminate the values of $V_1$ and $G_2$\,. 
 Two compact stars with different masses and radii, but with the same ratio $M/R$, 
have the same redshift. 
 For instance, this would be the case of two stars with the same redshift, one on the 
lower part of the $({\cal P}/{\cal E})_C$ curve and the other on the upper "crest", as 
we can observe in Fig.\,\ref{pipifig}\,.
 To avoid this ambiguity, we have attempted to explore the right hand side of 
Eq.\,(\ref{dpR}), by saying that even when the redshifts of two different stars are 
the same, their radii and/or masses are not.
 Unfortunately, the values of $[(dp/dr)/\varepsilon]_{r=R}$ are practically constant  
(around $-0.03741\,{\rm km}^{-1}$), with a variation of $\sim0.003\%$ within the 
parameter range, resulting inappropriate to our task. 
 So, in this first attempt to determine the parameters (based only on the values of M, 
R and $z_S$), we have obtained a large 
range of values for $V_1\in[0,\,0.5]$\,GeV and $G_2\in[0.0076,\,0.014]\,{\rm GeV}^4$, 
as we see in  the  $V_1$ vs. $G_2$ plot depicted by the solid line in Fig.\,\ref{1E1207a}.

\begin{figure}[th]
	\centering
        {\includegraphics[scale=0.4]{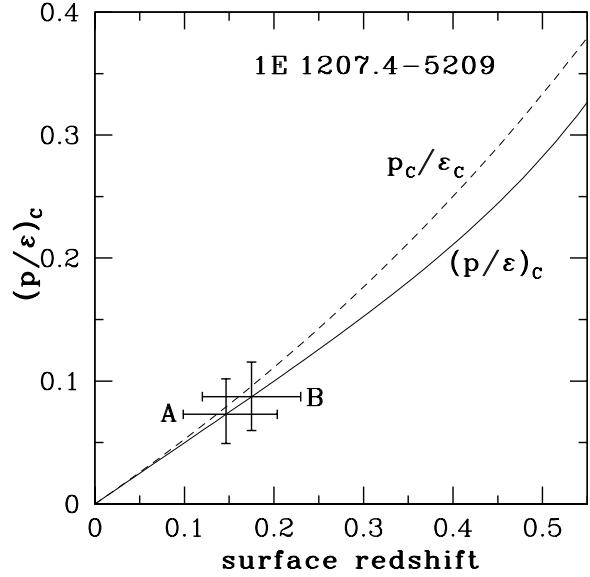}}
	\caption{ 
 As in Fig.\,\ref{pipifiga} but without the cloud of points. 
 The cross {\bf A} indicates the value of $({\cal P}/{\cal E})_C=0.073^{+0.029}_{-0.024}$ 
calculated by Eq.\,(\ref{pipi2}) corresponding to the redshift $z_S=0.15^{+0.057}_{-0.048}$\,.
  The cross {\bf B} indicates the value of $({\cal P}/{\cal E})_C=0.087\pm0.028$ calculated 
by Eq.\,(\ref{pipi2}) corresponding to the redshift $z_S=0.175\pm0.055$\,.
}
\label{1E1207}
\end{figure}

In order to narrow the search to get better results, we use the additional observable 
given by Eq.\,(\ref{pipi2})\,. 
 Briefly summarizing our strategy, we have generated $N$ random points (with normal 
distribution, which seems to be a reasonable assumption) within the error bars of the M 
and R determinations in Ref.\,\cite{Zha} 
to calculate the redshift values with which, in turn, through Eq.\,(\ref{pipi2}), we 
generate the corresponding points for $({\cal P}/{\cal E})_C$. 
 Then, applying the standard methods of data analysis in Refs.\,\cite{Bev1,Bev2}\,, 
we find $({\cal P}/{\cal E})_C=0.073^{+0.029}_{-0.024}$ at 68\% confidence (indicated by the 
cross {\bf A} in Fig.\,\ref{1E1207})\,. 
 As a result, within the same strategy (generating new random points within the error bars of 
$({\cal P}/{\cal E})_C$), we have obtained $V_1=0.44\pm0.11$\,GeV at 90\%  confidence
and $G_2=0.0082\pm0.001\,{\rm GeV}^4$\, at 95\% confidence (the cross {\bf A} in 
Fig.\,\ref{1E1207a}). 
 Then, according to these predictions (now taken at the central values of the model parameters), 
the compact star 1E\,1207.4-5209 is characterized 
by the central pressure ${\cal P}_C\simeq0.12\,{\rm GeV\,fm^{-3}}$ and energy density 
${\cal E}_C\simeq1.6\,{\rm GeV\,fm^{-3}}\simeq11\,\varepsilon_0$   
(where $\varepsilon_0\simeq0.141\,{\rm GeV\, fm^{-3}}$ is the nuclear energy density); 
the mass per baryon $m_C\simeq1.78\,{\rm GeV}\simeq1.89\,m_A$ at $r=0$ and 
$m_R\simeq1.75\,{\rm GeV}\simeq1.87\,m_A$  at $r=R$\,. 
 As a consequence of the high values of $V_1$ and $G_2$, the predicted 
anomalous mass defect is $|\Delta_2M|\simeq2.56\times10^{53}$\,erg\,.   

 Second, let us consider the the redshift range 0.12\,-\,0.23 from which we 
take the redshift $z_S=0.175\pm0.055$. 
 Making a similar calculation to that considered in the first step, we find 
the corresponding  observable $({\cal P}/{\cal E})_C=0.087\pm0.028$ at 71\% 
confidence  (indicated by the cross {\bf B} in Fig.\,\ref{1E1207} obtained 
from Eq.\,(\ref{pipi2}). 
 The dashed line in Fig.\,\ref{1E1207a} has the same meaning as the solid line in 
the first  step.
 By an analogous procedure to narrow our search we obtained $V_1=0.43\pm0.085$\,GeV 
at 94\% confidence and $G_2=0.0093\pm0.00092\,{\rm GeV}^4$\,at 94\% confidence.
 In this case, the compact star 1E\,1207.4-5209 is characterized by the central 
pressure ${\cal P}_C\simeq0.17\,{\rm GeV\,fm^{-3}}$ and energy density 
${\cal E}_C\simeq1.95\,{\rm GeV\,fm^{-3}}\simeq14\,\varepsilon_0$\,; 
the mass per baryon $m_C\simeq1.81\,{\rm GeV}\simeq1.93\,m_A$ at $r=0$ 
and $m_R\simeq1.77\,{\rm GeV}\simeq1.88\,m_A$  at $r=R$\,. 
 The corresponding anomalous mass defect is $|\Delta_2M|\simeq2.94\times10^{53}$\,erg\,.   
 We observe that the results for $|\Delta_2M|$ are not in contradiction with the 
ones shown in Fig.\,\ref{ponto3}\,. 
 The star 1E\,1207.4-5209 is neither at the maximum mass nor at the point {\bf 3} 
of the stellar sequence corresponding to the above values of $V_1$ and $G_2$.

 In both steps considered above, it is a remarkable feature of the FCM that the 
determination of the $Q{\bar Q}$ interaction potential and the gluon condensate 
from observations of the star 1E\,1207.4-5209 are in good agreement with $V_1=0.5$\,GeV, 
obtained from lattice calculations in Ref.\,\cite{KaZ}, and with 
$G_2=0.009\pm0.007\,{\rm GeV}^4$\, given by QCD sum rules calculations in 
Ref.\,\cite{Iof}, as shown in Fig.\,\ref{1E1207a}. 

\begin{figure}[th]
	\centering
	{\includegraphics[scale=0.4]{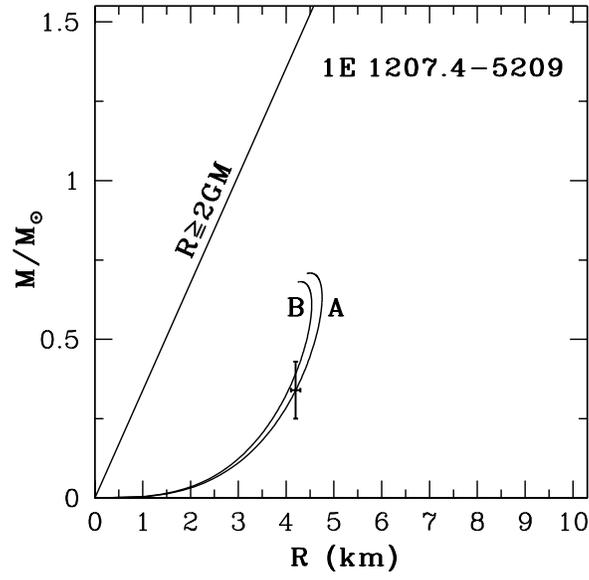}}
	\caption{
 The cross indicates the mass and radius of the star compact star 1E\,1207.4-5209. 
 The curves {\bf A} and {\bf B} are the mass-radius relations corresponding to the 
values of $V_1$ and $G_2$ indicated by the crosses {\bf A} and {\bf B} in  
Fig.\,\ref{1E1207a}, respectively.
}
\label{1E1207AB}
\end{figure}

 Fig.\,\ref{1E1207AB} shows the mass-radius relations up to the maximum masses 
corresponding to the two determinations of the parameters $V_1$ and $G_2$ as well 
as the location of the star 1E\,1207.4-5209 in the M-R diagram. 
 The curve {\bf B} is displaced to the left of the curve {\bf A} because of the 
error in the redshift assumed in the second step, which does not exactly satisfies   
Eq.\,(\ref{zs1}). 
 
\begin{figure}[th]
	\centering
	{\includegraphics[scale=0.4]{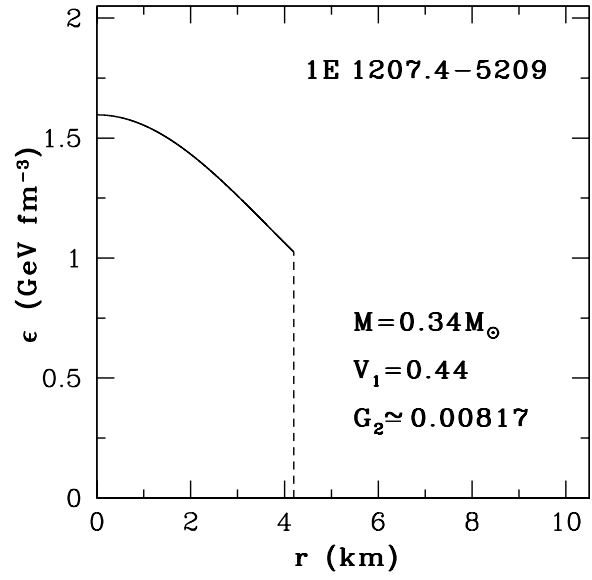}}
	\caption{
 Internal energy density profile of the compact star 1E\,1207.4-5209 
versus the radius $r$.
}
\label{1E1207ABa}
\end{figure}

 Another feature of the FCM is that, for increasing values of $V_1$ and $G_2$\,, the 
stellar configurations have lower masses and radii (cf. Fig. 2 in Ref.\,\cite{Fla3}). 
 In this case, it is not true that the Newtonian approximation is valid to calculate 
masses of strange stars, except in the very low mass regions, compared to the ones at 
the maximum masses (not the solar mass, as it is commonly accepted), of the stellar 
sequences.  
 The star 1E\,1207.4-5209 is a good example of a low mass compact star with a 
pronounced variation of the internal energy density profile from $r=0$ to $r=R$, 
as shown in Fig.\,\ref{1E1207ABa}\,. 
 

\section{Final remarks} 
\lb{fnlrmrks}

 In the present work, we have addressed the question of anomalous mass defects of 
low mass strange stars in the framework of the Field Correlator Method (FCM).
 The redshift measurements have played an important role in the determination of 
the model parameters $V_1$ and $G_2$ from astrophysical observations.
 In the case of the constant energy density solution of the Tolmann-Oppenheimer-Volkov 
equations, the ratio $({\cal P}/{\cal E})_C$ at the center of a compact star is an 
important observable quantity that can be determined from redshift measurements. 
 It tells us how the equation of state at $r=0$ is.

\begin{figure}[th]
	\centering
	{\includegraphics[scale=0.4]{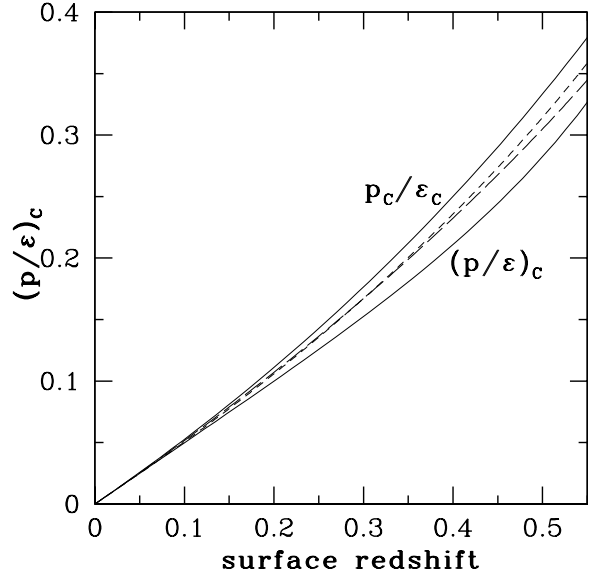}}
	\caption{ 
 Ratios pressure-to-energy density at $r=0$ as function of the redshift:  
 {\it Solid lines}\,, ${\cal P}_C/{\cal E}_C$ and $({\cal P}/{\cal E})_C$\, 
given by Eqs.\,(\ref{PezS}) and (\ref{pipi2})\,, respectively;
 {\it Short dashed line}\,, for the nuclear mean field theory in Refs.\,\cite{SW1,SW3}  
with the coupling constants  $(g_\sigma/m_\sigma)^2=11.798\,{\rm fm^2}$ and 
$(g_\omega/m_\omega)^2=8.653\,{\rm fm^2}$\, fixed to give the bind energy 
$E_{\rm b}=-15.75$\,MeV and $k_F=1.42\,{\rm fm}^{-1}$\,;
 {\it Long dashed line}\,, same as {\it Short dashed line}\, but for the arbitrarily chosen 
values: $(g_\sigma/m_\sigma)^2=15.0\,{\rm fm^2}$ and 
$(g_\omega/m_\omega)^2=12.0\,{\rm fm^2}$\,.
}
\label{mft}
\end{figure}

 In the general case, when the energy density is not constant, we have verified 
that the plots of $({\cal P}/{\cal E})_C$ vs. redshift, for different values of 
$V_1$ e $G_2$, are concentrated in a thin region with a quasi linear behavior, 
ranging from the origin to the maximum masses in the first branches of the stellar 
configurations. 
 This fact has enabled us to build a representation for the ratio $({\cal P}/{\cal E})_C$ 
in terms of the redshift similar to the one for the case of constant energy density.
 A remarkable feature of our approach is that the ratio $({\cal P}/{\cal E})_C$ as 
function of the redshift has the lowest values with respect to the models of nuclear 
matter.
 For instance, we illustrate in Fig.\,\ref{mft}\, the ratio $({\cal P}/{\cal E})_C$ for 
the Walecka nuclear mean field theory in Refs.\,\cite{SW1,SW3}. 
 Our preliminary calculations have shown that the values of 
$({\cal P}/{\cal E})_C$ $\forall\,z_S\in[0,\,0.5]$ are in the intermediate region between 
the solution for constant energy density and the one corresponding to the SQM in the FCM 
framework. 
 However, for different values of the coupling constants  $g_\sigma/m_\sigma$ and 
$g_\omega/m_\omega$ of the nuclear mean filed theory the curves of $({\cal P}/{\cal E})_C$ 
(in the redshift range we are considering) are not concentrated in a line of points as 
they are for the FCM. 
  An interesting task to be considered in future works is the investigation of the 
behavior of the ratio $({\cal P}/{\cal E})_C$ given by other models of nuclear matter in 
the framework of the mean field theories.
 Our attempt to determine ratio $({\cal P}/{\cal E})_C$ in terms of the 
redshift is model dependent, but with an almost model independent feature.
 It is important to verify if this feature remains valid for other approaches used to 
describe SQM, as the one considered in terms of the Richardson $Q{\bar Q}$ potential 
in Ref.\,\cite{SHS}. 
  
\centerline{\bf ACKNOWLEDGMENTS} 

This work was done with the support provided by the Minist\'erio da 
Ci\^encia , Tecnologia e Inova\c c\~ao (MCTI).
  

%
%

\newpage



\newpage


\newpage


\newpage


\newpage


\newpage


\newpage


\newpage


\newpage


\newpage


\newpage


\newpage


\newpage


\newpage


\newpage


\end{document}